# Applications of Earth-to-Air Heat Exchangers: A Holistic Review

Giouli Mihalakakou[a], Manolis Souliotis[b*], Maria Papadaki[a], George Halkos[c], John Paravantis[d], Sofoklis Makridis[a], Spiros Papaefthimiou[e]

[a] Department of Environmental Engineering, University of Patras, Greece

[b] Department of Chemical Engineering, University of Western Macedonia, Greece

[c] Department of Economics, University of Thessaly, Greece

[d] Department of International and European Studies, University of Piraeus, Greece

[e] School of Production Engineering and Management, Technical University of Crete, Greece

**Abstract**

The building sector is responsible for 40% of primary energy consumption, with heating/cooling covering the most significant portion. Thus, passive heating/cooling applications have gained significant ground during the last three decades, with many research activities on the subject. Among passive cooling/heating applications, ground cooling (especially earth-to-air heat exchangers) has been highlighted as a remarkably attractive technological research subjects because of its significant contribution to the reduction of heating/cooling energy loads; the improvement of indoor thermal comfort conditions; and the amelioration of the urban environment. This paper presents a holistic review of state-of-the-art research, methodologies, and technologies of earth-to-air heat exchangers that help achieve energy conservation and thermal comfort in the built environment. The review covers the critical subject of the thermal performance of earth-to-air heat exchanger systems; experimental studies and applications; parametric studies for investigating the impact of their main characteristics on thermal efficiency; and recent advances and trends including hybrid technologies and systems. The models describing the thermal performance of earth-to-air heat exchangers systems were classified in numerical, analytical, and data-driven; their main theoretical principles were presented; and experimental validation was mentioned when carried out. System parameters were grouped into three categories: system design, soil types, and soil surface coverage. System design parameters, especially length and burial depth, bore the most important influence on the thermal efficiency of the system. The paper was rounded up with an economic assessment of system application, and the conclusions highlighted the need for more experimental work including laboratory simulators.

*Corresponding author. E-mail address: msouliotis@uowm.gr (Manolis Souliotis)

**Contents**

# 1. Introduction

Buildings are responsible for approximately 40% of global energy consumption and for one-third of total energy greenhouse gas emissions [1]. The biggest part of the energy consumed in the building sector is used for space heating and cooling, which cover almost one third of total energy demand [2]. The use of conventional systems for building heating/cooling requirements present a significant energy and environmental impact, such as increase of carbon dioxide ($CO_2$) emissions; global warming; aggravation of greenhouse and urban heat island effects; increase of peak electric load demands; degradation of indoor air quality, etc. [3]. During the last decades, the scientific and technical community has made a great effort to explore and investigate the most energy efficient solutions, based on renewable energy sources (RES), for space heating/cooling which could simultaneously contribute to energy conservation and environmental protection [4]–[6]. A remarkable place among these solutions and measures is held by passive heating and cooling techniques and algorithms, which could offer a significant reduction of energy consumption, resulting in a potentially remarkable mitigation of the heat island effect and improvement of the urban microclimate [7].

Passive cooling/heating presented an impressive evolution during the last three decades, enriched with new research, techniques, and materials, and it can provide minimization of heating/cooling loads; excellent thermal comfort conditions; mitigation of the urban heat island; and an amelioration of the urban environment [8]–[10]. Passive heating/cooling techniques are mainly based on the three natural heat transfer mechanisms: conduction, convection and radiation while passive cooling uses extensively the following three natural heat sinks for heat dissipation: ground, sky, and water. The ground can act as a heat sink for cooling a building or an agricultural greenhouse, but also as a heat source for providing heating during the cold periods of a year. Ground cooling/heating is one of the most documented and widely applied passive methods based mainly on thermal properties and temperature distribution at the surface of the ground as well as below it, as ground temperature remains constant at a depth of 2.5 to 3m throughout the year [11], [12]. Ground cooling/heating includes the following two main strategies: (a) direct earth-coupling techniques with earth-sheltered buildings [13], [14]; and (b) indirect earth-coupling techniques with earth-to-air heat exchangers [15]–[17].

Earth-to-air heat exchangers (EAHE) use ground as a heat sink/source and consist of a single or multiple pipes buried in the ground, through which ambient or indoor air is circulated and heat is transferred from the air to the soil during the summer (cooling mode) or from the soil to the air during the winter (heating mode). The air at the outlet of the pipes is mixed with the indoor air of a building or an agricultural greenhouse [11], [18]–[20]. The thermal performance of EAHE is strongly influenced by many parameters, which could be classified in three main categories: (a) system design parameters, such as pipe material, pipe length, pipe radius, burial depth, and number of pipes; (b) different soil types described and expressed by the thermophysical properties of the soil profile, such as moisture content, thermal conductivity, specific heat capacity, and thermal diffusivity; (c) environmental parameters responsible for the temperature distribution of the ground surface, such as short and long-wave radiation, convective heat transfer, wind speed, vegetation and ground surface coverage, and evaporation/condensation speed [1], [18]. Figure 1 represents a single EAHE including the most important environmental and heat transfer processes influencing the thermal performance of the system; these may be summarized as follows: (a) environmental and thermophysical processes responsible for the temperature distribution at the surface and various depths of the ground, such as short and long wave radiation, convective heat flux, ground surface coverage, evaporation, and wind speed; and (b) heat transfer mechanisms in the soil and pipe, including conduction in the soil and convection inside the pipe.

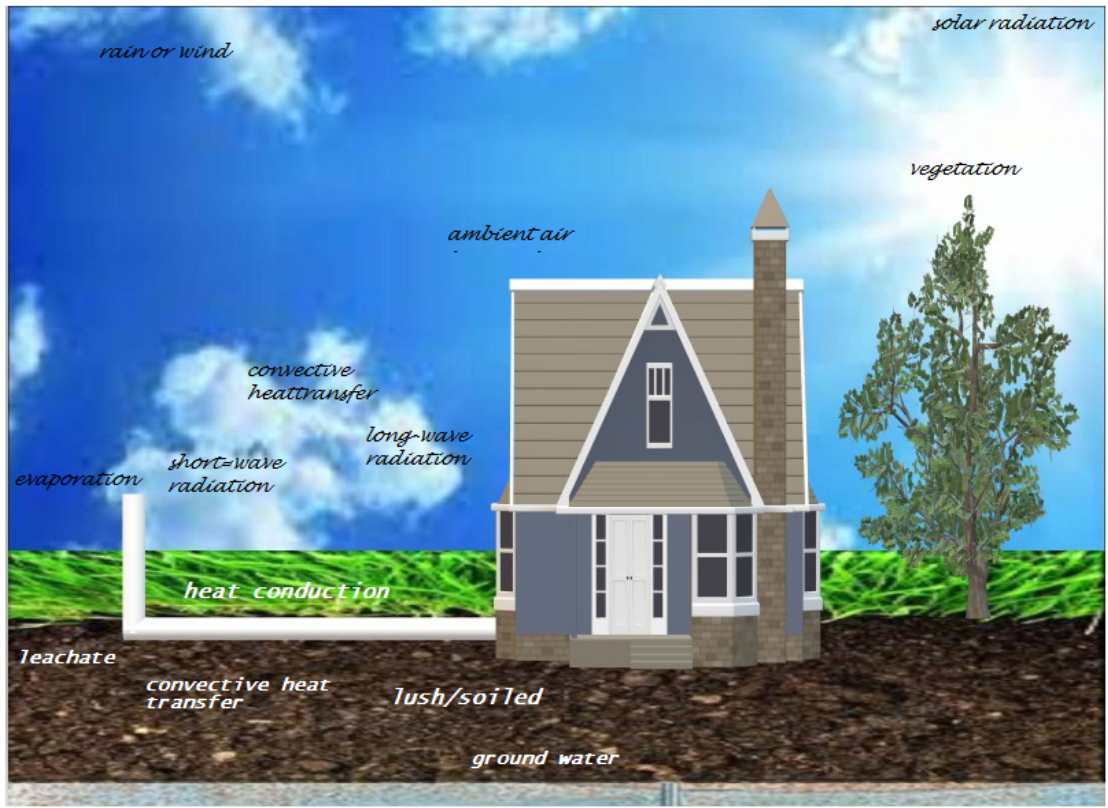

**Figure 1.** EAHE and main environmental and heat transfer mechanisms affecting system performance.

This paper aims to offer a holistic and critical perspective on the subject of EAHE systems by presenting a complete and integrated review of the following main research activities: (a) modeling, (b) experimental investigation, (c) parametric studies, (d) hybrid systems, and (e) economic assessment. Modeling the thermal performance of EAHE systems is an attractive subject encompassing many mainly deterministic models and a few data-driven ones. Deterministic models, which could be numerical [16], [21]–[48] or analytical, [49]–[59],[60], are mainly based on heat and mass transfer processes in the air and in the soil, while data-driven models use historical data to train and test artificial intelligence networks [61]–[63]. Moreover, a significant number of research studies describe experimental investigations and case studies of different EAHE systems, including open and closed systems, various climatic conditions, tube shapes, soil types and coverage. In addition, experiments have been carried out and presented to validate mathematical models, present innovative experimental configurations, and (mainly) estimate and assess the heating or cooling potential of the system [16], [17], [64]–[69].

Apart from modeling and experimental studies, one of the most interesting topics of EAHE research has been the investigation of the impact of main system design parameters (such as pipe material, pipe length, pipe diameter, burial depth, air speed inside the tube, and number of pipes) on the heating/cooling efficiency of the system. Another interesting topic has been the extensive study and analysis of a significant number of environmental parameters (such as soil type, and soil coverage) on the capacity of the system [36], [70]–[74].

This paper also presents a review of recent EAHE system advancements, including combined and hybrid technologies to improve system efficiency, to minimize energy consumption, and provide good thermal comfort conditions. These hybrid technologies could include photovoltaic systems, integrated solar chimneys, solar and wind systems, coupled ventilated roofs, hydrogen technologies, buried pipes filled with phase change materials, etc. [1], [75], [84]–[93], [76]–[83]. The presentation is rounded up with an economic assessment included in order to complete a holistic approach to the subject. The rest of this manuscript is structured as follows. Section 2 covers modeling, with Section 2.1 focusing on numerical, Section 2.2 on analytical, and Section 2.3 on data driven models. Section 3 covers experimental while Section 4 covers parametric studies. Section 5 covers hybrid systems and Section 6 reviews economic studies. Finally, Section 6 presents the conclusions.

## 2. Modeling the thermal performance of earth to air heat exchangers

The ground temperature distribution and the energy balance at the ground surface are significant parameters for passive cooling/heating and especially for direct (earth-sheltered buildings) and indirect (EAHE systems) ground cooling/heating. The ground temperature at the surface and various depths within the soil profile has been investigated by both experimental and theoretical methods [13], [15], [53], [79], [94]–[107]. Modeling and assessing the thermal performance of an earth-to-air heat exchangers system requires a deep knowledge of ground temperature profiles. That means knowledge of daily and annual fluctuations of soil temperature at the surface of the ground and various depths below it. Taking into account mainly the crucial parameter of ground temperature distribution at the ground surface and various depths below it, many models for simulating and predicting the EAHE systems have been developed and presented in the scientific literature [21]–[32], [42]–[48]. Santamouris and Kolokotsa [7] classified models in (a) numerical and analytical deterministic models, where the entire problem is described and solved through a mathematical formulation, and (b) data-driven models, where predictions were achieved through training the proposed model with the appropriate set of historical data. In this research, the EAHE models are classified into three categories: (a) numerical [16], [21]–[48], [73], [74], [108], (b) analytical [49]–[52], [54]–[59], [109], and (c) data-driven [61]–[63].

## 2.1. Modeling earth to air heat exchangers using numerical methods

Many numerical models have been developed and proposed during the last few decades for simulating and predicting the thermal performance of an EAHE system for heating/cooling of a building or of an agricultural greenhouse. These models are based on differential equations describing mainly the conductional and convectional heat transfer processes in the soil and inside the tubes respectively during the operation of the

EAHE. The large majority of that type of models have adopted several approximations (e.g. geometries, initial and boundary conditions, etc.) for discretizing and solving the differential equation sets. According to [21], numerical models describing the thermal behavior of EAHE systems could be classified into two categories: (a) models where only a part of the proposed geometry is influenced by the pipe presence; and (b) models where the entire proposed geometry is considered influenced.

Numerical models are mainly based on conductional processes in the soil, caused by: (a) the ground surface boundary conditions described by the energy balance equation at the surface; and (b) the EAHE system's presence, as well as convection phenomena inside the tube. Some of the models which focus on a full description of heat transfer processes in the soil, mainly take into consideration only conductional processes. These models consider specific boundary conditions including the ground surface and the undisturbed ground and the pipe, while they also describe the convectional processes necessary for calculating the air temperature distribution inside the tube. Moreover, there are some numerical models which are focused on heat convectional phenomena regarding the air flow inside the tube, and employ computational fluid dynamics theory and modeling.

The following numerical models may be classified in the first category, including models with a complete and detailed description of heat transfer processes in the soil. Puri [32] proposed a transient model of a set of two differential equations expressing the coupled and simultaneous heat and moisture transfer in the soil for a single EAHE. He selected a one-dimensional axial geometry considering that soil temperature and moisture around the tube were constant initially, while at the boundary of a large axial distance, ground temperature distribution and moisture content were not influenced by the tube's presence. Soil and air parameters such as soil thermal conductivity and diffusivity were taken into account and the system of the two differential equations with two independent variables (axial distance and time) was discretized numerically using the finite elements method. Those authors found that soil temperature profiles evolve at a faster rate than the corresponding moisture content.

The energy and mass balance equations of [32] were the basis for the development of a transient numerical model developed by [42]. This model considers one single EAHE, takes into account the simultaneous heat and mass transfer into the soil and pipe, and it is expressed in polar co-ordinates with three independent variables (axial distance, y-co-ordinate, and time) and two dependent (heat and moisture). Many thermophysical soil and pipe parameters were taken into account, while boundary conditions were set axially at a large distance from the pipe, where it was assumed that temperature and moisture profiles are not influenced by the pipe's presence and at the pipe vicinity, where heat transfer depends on heat losses from the pipe to the ground and moisture migration due to the exchanger is zero with the pipe considered impervious. The differential equations were discretized and solved using the numerical method of control-volume and the results were validated and found sufficiently accurate against extensive sets of experimental data. The model was extended to calculate the thermal performance of a system of multiple EAHEs using the mathematical algorithm of superposition [43].

Bojic et al. [44] developed a set of linear equations describing the thermal performance of an EAHE system for heating/cooling a building. The mathematical model of the EAHE system consists of steady-state heat balance equations applied in each soil layer, assuming that soil is divided in layers with uniform temperature values. Pipes are considered all on the same layer. EAHE system was coupled with a building and authors calculated its energy needs and the contribution of the EAHE system to these needs. As shown, the system, with its main configuration parameter values, can effectively cover a significant part of building's energy needs especially during summer.

A transient numerical model simulating the thermal behavior of an EAHE system used for energy conservation in an agricultural greenhouse was developed in [45]. That model was based on heat transfer differential equations describing the energy movement into the soil in three dimensions (x, y, z). A number of assumptions were considered mostly regarding soil thermal properties. The soil moisture gradient was considered not to affect the temperature gradient while the model took into account thermal processes and evaporation/condensation inside pipes. The numerical method of finite difference was selected for discretization and the model was successfully validated against experimental data.

A mathematical model combining numerical and analytical solutions was presented in [46]. The model calculated the soil and air temperature distributions using the form of Fourier integrals that correspond to an EAHE system being used for ventilation purposes. One-dimensional heat conduction and convection equations were used for calculating soil and air temperature inside the tube respectively. The ground temperature profile was estimated as a superposition of the undisturbed temperature profile and the temperature profile caused by the presence of the exchanger. The model was validated experimentally.

A two-dimensional set of heat and mass transfer differential equations describing thermal processes during the operation of a single earth tube for space heating, was solved numerically in [47]. Boundary conditions were specified as the undisturbed ground temperature at a large distance from the pipe and as the energy balance equation at the ground surface, including an extensive number of parameters such as solar radiation, long wave radiation, convective and latent heat fluxes at ground surface, etc. All used parameters were defined while thermal balance equations and convection theory were used for estimating the air temperature distribution inside the pipe. The proposed model was developed and integrated inside an existing computing environment developed to simulate passive systems for heating in buildings. A parametric study was also presented.

Yoon et al. [48] presented a transient numerical method based on heat conduction equations in the soil, and taking into account the energy balance equation at the soil surface as well as the pipe presence for describing the thermal behavior of a system with multiple buried pipes used for space heating/cooling. A complete design description was offered for the multiple pipes and results were validated against measurements. Tittelain et al. [21] proposed a transient numerical model for calculating the thermal performance of buried pipes. That model was based on discretizing the conductive flow using a response factor method. As regards geometry, a perpendicular to the pipes n-sections method was used to discretize the system. Validation was achieved by comparing theoretical results with those of two other experimentally validated models, one analytical and one numerical.

The computing environment of the EnergyPlus simulation program (https://energyplus.net/) was used to solve numerically the heat and mass transfer equations describing the thermal behavior of an EAHE system presented in [22]. The model took into account many parameters influencing the ground surface temperature profile, such as solar radiation, convection between ground and air, latent heat fluxes, etc. The results were successfully validated against measured values and the system proved efficient for energy conservation, able to provide 86% of the cooling load of July.

Trzaski and Zawada [23] proposed a three-dimensional, finite element numerical model for heat and mass transfer processes in soil and pipes, including an extensive number of parameters influencing the thermal behavior of the system, especially the air temperature at the pipe outlet. Those parameters were divided into three main categories: (a) system geometry (pipe length, burial depth, number of pipes, etc.); (b) condition of the ground surface (vegetation, ground shading, solar radiation, etc.); and (c) thermophysical parameters of the ground (density, moisture content, specific heat capacity, etc.). An extensive number of simulations were performed for different values of the aforementioned parameters, and their influence on the system efficiency was investigated. The results of the model were validated successfully against experimental data. A transient, three-dimensional numerical model was also presented in [24], taking into account the coupled and simultaneous heat and mass transfer in the soil and pipe, during the performance of an EAHE system for cooling and pre-heating indoor spaces. A set of differential equations describing the mass and heat transfer in the soil was discretized and solved numerically using a control volume formulation. Boundary conditions were described analytically in all dimensions. The model results were compared successfully with the results of another model that was experimentally validated [110].

A two-dimensional, pseudo-transient numerical model describing the thermal performance of an EAHE system for space heating/cooling in three cities in Mexico, was developed and presented in [25]. The model was based on energy, mass, and momentum conservation equations, which were discretized and solved numerically with finite elements, using the restrictions of specific boundary conditions at the ground surface, at the pipe, and at a significant depth. Four Reynolds numbers were used for the simulations (100, 500, 1000, 1500) and the results showed remarkable potential for the heating/cooling system. Another two-dimensional, transient, numerical model was presented in [26], based on a combination of conductional and convectional heat transfer processes and using a control volume formulation for discretization. The profile estimation of the soil temperature was based on conduction theory, and the soil was divided in control volumes while boundary conditions were set at the soil surface, at the undisturbed ground, and at the tube. The convectional heat transfer process was taken into account for the air flow inside the tube, and Nusselt numbers were calculated for heating and cooling mode. Theoretical results were validated against existing experimental data and the model was found sufficiently accurate.

Fazlikhani et al. [27] and Belatrache et al. [28] offered methodologies for predicting the thermal performance and efficiency of EAHE systems in cold and hot arid climates in Iran and in arid climates in Algeria, respectively. Those methodologies were based on both conductive heat transfer in soil, formulating the soil temperature distribution at the surface and at various depths below it, and the convective heat transfer for the air flow inside the tube, thus calculating the air temperature inside the pipe. Both models were based on heat transfer theory, while differential equations were solved numerically and results were successfully validated against experimental data.

Rodrigues et al. [29] developed a transient, numerical model simulating the thermal performance of an EAHE system for different soil types, and for three different places in a coastal area of Brazil. Heat conduction equations were used for the soil temperature distribution, while the set of differential equations describing continuity, momentum, and energy was considered for the simulation of the air flow inside the tube, thus for the estimation of the air temperature inside the tube.

Liu et al. [30] developed a numerical model for predicting the thermal performance of a vertical EAHE system. Conductional and convectional heat transfer equations were used for estimating the distributions of the soil and the air temperature. Three models were developed for air, tube, and soil, while a control volume formulation was used for the discretization process. Code was developed in MATLAB, while the results were compared with experimental data and the model was found sufficiently accurate.

A numerical, transient, two-dimensional model based on heat exchanged between the air flowing inside the tube and the soil was developed in [74]. That model took into account an extensive number of soil thermal properties, and it was mainly used to investigate the impact of different soil coatings on the thermal performance of the system. The model results were validated experimentally.

Attention now turns to research studies that included numerical models mainly based on convectional processes for different air flow patterns inside a tube. The most common characteristic of these models is the extensive use of computational fluid dynamics (CFD) simulation software. Hollmuller and Lachal [31] developed a numerical model based mainly on heat and mass transfer processes between air and tube, which takes into account both sensible and latent heat fluxes and exchanges. The tubes were divided into nodes from inlet to outlet, and sensible heat was calculated using convective heat transfer equation, while latent heat was determined using water transfer equations for evaporation or condensation. Calculations of energy and water balance were performed for all tube nodes, and the heat transfer to soil nodes was calculated afterwards. Numerical results were extensively validated in four experimental sites, and the model was found sufficiently accurate.

The PHOENICS software [111] was used for the development of a transient, three-dimensional numerical model describing the thermal performance of an EAHE system in [33]. The numerical model used implicit methods based on simultaneous heat and mass transfer processes in the ground and pipe during the operation of the system, and it combined heat and mass transfer in the soil with convective fluid dynamic of turbulent flow inside the tube. Model results were tested and found to be in close agreement with experimental data.

CFD modeling, to predict and estimate the air flow and temperature variation inside the tubes using numerical methods was used in [34]. That research was focused on fluid dynamics inside the duct, based on convective heat transfer theory for turbulent flow. Model results were validated against measured data while an artificial neural network (ANN) model was designed and trained for the estimation of convective heat transfer coefficients such as Nusselt numbers. Neural model results were compared and tested with those of the CFD model and found to be in close agreement.

Vaz et al. [16] offered a numerical solution of a set of differential equations consisting of (a) the Navier-Stokes equations describing the fluid mass flow conservation inside the tube; and (b) energy conservation equation as fluid also creates heat transfer inside the tube, thus calculating the temperature at each point of the tube. Model results were validated experimentally and the model was found to be accurate. Su et al. [35] developed two numerical sub-models for the description of a deeply buried EAHE system for space heating/cooling. The first one calculated air temperature values and moisture content inside the pipe, and was expressed by one-dimensional convective heat and mass transfer differential equations. The equations were discretized and solved numerically and implicitly, thus calculating the air temperature and moisture content at each of the defined node and finally at the pipe exit. The second sub-model described the ground temperature distribution by solving numerically the one-dimensional transient heat conduction equation in the soil, taking into account a large number of soil thermophysical parameters and using radial co-ordinate. Model results were tested with measurements and found to be in close agreement.

Sehli et al. [36] proposed a numerical, one-dimensional model to predict the thermal performance of an EAHE system for space heating/cooling. The model was mainly based on convective heat and momentum differential equation for turbulent flow, to estimate and predict the air temperature values at every point inside the tube. The convective parameters of Reynolds number and form factor were extensively used to simulate the system performance. The analytical solution of heat conduction equation at the ground surface and at various depths were used for the ground boundary conditions. Model results were successfully validated against experimental data.

A transient numerical model based on fluid mechanics with optimized geometry was developed by [37] to predict the thermal behavior of an EAHE system. Modeling used the convective flow differential equations for energy, mass, and momentum, while the one-dimensional conductive heat transfer equation in the soil was used for estimating the soil temperature profile. Model results were found to be in close agreement with literature data. Fluid dynamics modeling was also used to estimate the thermal performance of EAHE in the winter season for the hot and dry climates of India in [38]. The CFD model was based on the numerical solutions of three-dimensional Cartesian-coordinate differential equations expressing energy and mass transfer during the performance of an EAHE system. The set of continuity, momentum, and energy equations was solved numerically for specific boundary conditions especially at the tube surface and at the soil-pipe interface, while air temperature values were calculated at any point of the tube. Model results were successfully validated against experimental data.

A transient, numerical, three-dimensional simulation program based on CFD modeling was developed and presented in [73] for estimating the EAHE thermal performance for space cooling. The model was based on fluid mechanics theory and analysis, considering the fluid flow governing the differential equations of continuity, heat, and mass transfer. The model was solved numerically in the FLUENT environment, and its results were successfully validated against experimental data.

Ramirez-Davila et al. [39] developed a numerical model based on fluid mechanics to calculate the thermal performance of EAHE for three cities in Mexico. A finite volume formulation was used to discretize the three-dimensional differential equations describing energy, continuity, and momentum inside the pipe and the ground. Boundary conditions were set at the ground surface, where a great number of involved parameters were considered at the undisturbed ground temperature and at the pipe. For air temperature at the pipe inlet and outlet, convective heat transfer flow and computational fluid dynamics were used. Discretization was achieved using a CFD code and a finite elements methodology to calculate temperature values at any point inside the tube and at the exit, as a function of Reynolds number. The model was validated against measurements, and it was concluded that the system could be efficient for building heating and cooling.

Momentum, continuity, and energy differential equations were used in [26] as the governing equations for describing the thermal behavior of an EAHE system. The equations were solved numerically using a computational fluid dynamic piece of software and the finite volume method. That work also investigated the soil temperature recovery caused by the EAHE performance. The model results were successfully validated against experimental data.

Serageldin et al. [41] developed a transient mathematical formulation for predicting the thermal performance of an EAHE system for space heating/cooling and for the climate of Egypt. The model was based on a one-dimensional energy conservation equation, with a great number of assumptions concerning mainly the soil thermal condition and properties and the air flow inside the pipe. A convectional three-dimensional model based on a CFD simulation software was also used to predict the air flow and temperature inside the pipe. The model results were extensively validated against experimental data.

Ahmed et al. [108] offered a CFD, two-dimensional, numerical model for describing the thermal performance of a horizontal EAHE system for space cooling. The model was based on the Navier-Stokes equations for momentum conservation, and on energy conservation equations, and it used the FLUENT simulation software [112] for solving the differential equations numerically. The model results were successfully validated against experimental data.

## 2.2. Modeling earth-to-air heat exchangers using analytical methods

Apart from numerical models, an extensive set of analytical models have been proposed in scientific literature. A simple, one-dimensional analytical model based on heat transferred inside a tube was developed and presented in [49] for predicting the thermal behavior of an earth-air tunnel for heating/cooling a hospital building in India. The model was validated against experimental data and found to be sufficiently accurate.

Santamouris and Lefas [50] developed a one-dimensional analytical model for predicting the thermal behavior of an EAHE system for a hybrid agricultural greenhouse. The model was based on steady-state energy balance equations. The model was validated against existing measurements.

An analytical parametric model was implemented by [52] for estimating the thermal behavior of an EAHE system. The model was based on a systematic analytical parametric process developed and expressed using regression analysis. The following four parameters, influencing the thermal performance of the system, were used for the analysis: (a) pipe length; (b) pipe radius; (c) air velocity inside the tube; and (d) burial depth below the ground surface. The model results were successfully validated against accurate numerical data as well as experimental data, and the model was found accurate.

Krarti and Kreider [109] presented a simplified, transient analytical model for predicting EAHE thermal performance. The model, which was able to estimate the air temperature at the tube exit at any time, was based on heat transfer equations for conduction and convection, and it assumed that the system arrives at a quasi-steady state operation after a few days of performance. The model was successfully validated against measurements. A transient, analytical, three-dimensional model for predicting the thermal performance of an EAHE system was developed and presented in [54]. The model was based on the combination of heat flux and steady state equations. Authors did not offer experimental validation.

Deglin et al. [60] developed an analytical three-dimensional, transient model for predicting the thermal performance of a subsoil EAHE system for air-conditioning of a livestock buildings. The model was mainly based on convectional heat transfer processes between the air inside the tubes and the soil. The theoretical results were successfully validated against experimental data.

An analytical, steady-state, one-dimensional model was presented in [55] for predicting the air temperature distribution at the pipe exit of an EAHE system used for cooling a residential building in a desert climate. The model was based on calculating thermal resistances due to steady state thermal analysis and to convective flow inside the tube. The convective heat transfer coefficient was calculated as a function of Reynolds and Nusselt numbers for laminar and turbulent flow. The model was successfully validated against experimental and theoretical parametric studies. Cucumo et al. [56] developed a transient, one-dimensional, analytical model for predicting the thermal behavior of an EAHE system. The model was based on heat conductional fluxes for a semi-infinite body as well as convective heat transfer, taking into account the latent heat flux inside the pipe. The model results were successfully validated against experimental data.

Lee et al. [57] offered an analytical transient module, developed inside the environment of the EnergyPlus simulation program, based on soil temperature distribution and thermal resistance equations due to convection phenomena caused by the tube presence. The model results were successfully validated against experimental and theoretical data. Do et al. [58] implemented an analytical, one-dimensional, steady state model to simulate a closed-loop EAHE for space cooling of residential buildings in hot and humid climates. Their model was simple and based on the analytical solution of conductive heat transfer equations in the soil as well as on thermal resistances of system components. The model was successfully validated against experimental data from published studies.

A transient, analytical approach describing the thermal behavior of an EAHE system was developed by [59]. The model was based on the harmonic temperature fluctuations of system components and of its thermal environments caused by conduction and convection processes. The model results were successfully validated against theoretical data received from validated numerical models. Rouag et al. [51] implemented a transient, one-dimensional, semi-analytical model for simulating the soil temperature profile around an EAHE system. That model was based on heat conduction processes in the soil in the pipe's vicinity. The model was successfully validated against literature results.

## 2.3. Modeling earth to air heat exchangers using data driven models

Data driven models are computing systems based on historical data instead of the mathematical and physical formulations and equations of the deterministic models. Artificial Neural Networks (ANNs) models, which simulate biological neurons, and fuzzy logic models belong to this category of data driven models. The most important characteristic of the data driven models is their efficiency, in the sense that the analysis depends only on the available historical data, allowing the modeling of rather complicated non-linear systems and processes with sufficient accuracy. Therefore, they often present significant advantages compared with deterministic models, including their simplicity as all they need is training and testing with historical data. Data driven models should be well trained to achieve an acceptable level of error between target and testing set of data, and then the system is ready for use.

Mihalakakou [61] presented an ANN approach for estimating the air temperature at the tube exit. The model was multiple layered, based on a backpropagation algorithm, and the input set of data included the following parameters: (a) air temperature; (b) relative humidity; (c) ground surface temperature; and (d) ground temperature at burial depth, while the output parameter was the air temperature at the pipe exit. The model was extensively trained and tested and its results were successfully validated against the results of an accurate deterministic model, proving its efficiency in predicting the thermal behavior of an EAHE system.

A neural network model designed, trained, and tested to estimate air temperature distribution at the EAHE exit was presented in [62]. The neural network model had a typical multi-layered feed-forward architecture based on the back-propagation method, while a significant number of input parameters were selected. The results were extensively tested, compared with those of an experimentally validated numerical model [42], and the neural model was found accurate and appropriate for use. Diaz et al. [63] developed both a thermodynamic model simulating the thermal performance of an EAHE system, and a fuzzy logic instead of an on/off controller, which could be coupled with the system in order to improve its energy efficiency. Their results showed that the use of the fuzzy logic controller could reduce energy consumption for heating/cooling of a building by almost 88%. A summary of the characteristics of the aforementioned models simulating the thermal performance of an EAHE system, numerical, analytical, and data driven is displayed in Table 1.

**Table 1.** Main characteristics of considered models simulating the thermal performance of an EAHE system

| NUMERICAL MODELS | | |
|---|---|---|
| **Reference** | **Short Description** | **Validation Mode** |
| [32] | Transient, based on coupled and simultaneous heat and moisture transfer in the soil. | Author did not mention any validation process |
| [42] | Transient, for a single pipe with three independent variables, based on coupled and simultaneous heat and moisture transfer into the soil caused by the tube's presence | Theoretical results were compared with experimental data and model was found accurate |
| [43] | Transient, for multiple pipes, based on coupled and simultaneous heat and moisture transfer into the soil, it uses the mathematical method of superposition for multiple pipes | Model was successfully validated against an extensive set of experimental data and found accurate |
| [44] | Steady state heat balance equations. | Authors did not mention any validation process |
| [45] | Transient, three dimensional, based on heat transfer differential equations describing the energy movement into the soil in three dimensions (x,y,z) | Model was successfully validated against experimental data and analytical predictions |
| [46] | Transient, based on representation of temperature distribution in the form of Fourier integrals. | Model was experimentally validated |
| [47] | Transient, two-dimensional, focused on heat conductional processes into the soil | Author did not mention any validation process |
| [21] | Transient, based on discretizing the conductive flow using a response factor method | Validation was achieved by comparing results with those of two other experimentally validated models, one analytical and one numerical |
| [48] | Transient, based on heat conduction equation in soil and taking into account the energy balance equation at the soil surface | Theoretical results were compared with experimental data and the model was found accurate |
| [22] | Transient, takes into account a great number of parameters influencing the ground surface temperature profile such as solar radiation, convection between ground and air, latent heat fluxes, etc. | Model results were successfully validated against measured values and the system was proved efficient for energy conservation, able of providing the 86% of cooling load in July. |
| [23] | Three-dimensional, finite element numerical model describing heat and mass transfer in soil and pipes including an extensive number of involving parameters influencing the thermal behavior of the system | Theoretical results were compared with experimental data and the model was found accurate |
| [24] | Transient, three-dimensional, based on the coupled and simultaneous heat and mass transfer in the soil and pipe | Model results were validated against the results of an experimentally validated model |
| [25] | Pseudo transient, two-dimensional, based on energy, mass, and momentum conservation equations | Authors did not mention any validation process |
| [26] | Transient, two-dimensional, based on a combination of conduction and convectional heat transfer | Theoretical results were validated against existing experimental data and the model was found accurate |
| [27] | Transient, one-dimensional, based on both conductive heat transfer in the soil and on convective heat transfer in the air flow inside the tube | Model results were successfully validated against experimental data. |
| [28] | Transient, one-dimensional, based on both conductive heat transfer in the soil and on convective heat transfer in the air flow inside the tube | Model results were successfully validated against experimental data. |
| [29] | Transient, one-dimensional, based on conductional processes in the soil and on the set of differential equations describing continuity, momentum and energy equations for air flow inside the tube | Authors did not mention any validation process |
| [30] | Transient, based on conductional and convectional heat transfer equations used for estimating both soil and air temperature distribution | Model results were compared with experimental data and the model was found accurate |
| [31] | Steady state, based on sensible and latent heat fluxes and exchanges | The model was successfully validated against an extensive set of experimental data and it was found accurate |
| [33] | Transient, three-dimensional, based on simultaneous heat and mass transfer, focused on convective processes in the pipe | Model results were tested against experimental data and were in close agreement |
| [34] | Focused on fluid dynamics inside the duct and based on the convective heat transfer theory for turbulent flow | Model results were successfully validated against measured data. An artificial neural network model was also designed and trained for the estimation of convective heat transfer coefficients. Neural model results were compared and tested with those of the CFD model and they were in close agreement. |
| [16] | Transient, based on the Navier-Stokes equations describing the fluid mass flow conservation and on the energy conservation equation as the fluid flows inside the tube | Model results were validated experimentally and the model was found accurate |
| [35] | Transient, consists of two sub-models describing heat conduction and convection processes into the soil | Model results were tested with measurements and they were in close agreement. |
| [36] | One-dimensional, based on convective heat and momentum differential equation for turbulent flow | Model results were successfully validated against experimental data |

| Reference | Short Description | Validation Mode |
|---|---|---|
| [37] | Transient, based on the convective flow differential equations for energy, mass, and momentum | Model results were compared with literature data and they were in close agreement. |
| [38] | Transient, three-dimensional based on computational fluid dynamics modeling | Model results were validated against experimental data and the model was found accurate. |
| [39] | Two-dimensional, based on fluid mechanics and on energy, continuity, and momentum inside the pipe and the ground | Model was validated against measurements and it was concluded that the system can be efficient for building heating and cooling. |
| [26] | Transient, based on computational fluid dynamics modeling | Model results were successfully validated against experimental data |
| [41] | Transient, one-dimensional, based on computational fluid dynamics modeling | Model results were successfully validated against experimental data |
| [108] | Transient, two-dimensional, based on computational fluid dynamics modeling | Model results were successfully validated against experimental data |
| [73] | Transient, three-dimensional, based on computational fluid dynamics modeling | Model results were successfully validated against experimental data |
| [74] | Transient, two-dimensional, based on heat exchanged between the air flowing inside the pipe and the soil | Model results were successfully validated against experimental data |
| **ANALYTICAL MODELS** | | |
| **Reference** | **Short Description** | **Validation Mode** |
| [49] | Steady state, one-dimensional | The model was validated against experimental data and found accurate. |
| [50] | Steady state, one-dimensional | The model was successfully validated against existing measurements. |
| [52] | Parametric model, regression analysis | Model results were successfully validated against accurate numerical data as well as against experimental data and the model was found accurate |
| [109] | Simplified, transient | Model results were successfully validated against measurements |
| [54] | Transient, three-dimensional | Authors did not mention any validation process |
| [55] | Steady state, one-dimensional | Model results were successfully validated against experimental and theoretical parametric studies |
| [56] | Transient, one-dimensional | Model results were successfully validated against experimental data |
| [57] | Transient, inside the EnergyPlus environment | Model results were validated against experimental and theoretical data and found accurate. |

| Reference | Short Description | Validation Mode |
|---|---|---|
| [58] | Steady state, one-dimensional | The model was successfully validated against experimental data from published experimental studies |
| [59] | Transient, based on harmonic temperature fluctuations of system components. | Model results were successfully validated against simulation data from validated numerical models |
| [51] | Transient, one-dimensional, semi-analytical | The model was successfully validated against literature results |
| [60] | Transient, three-dimensional | The model was successfully validated against experimental data |
| **DATA DRIVEN MODELS** | | |
| **Reference** | **Short Description** | **Validation Mode** |
| [61] | Artificial Neural Network model based on back-propagation algorithm | The model was extensively trained and tested and its results were successfully validated against the results of an accurate deterministic model |
| [62] | Artificial Neural Network model based on back-propagation algorithm | Results were extensively tested and compared with those of an experimentally validated numerical model and the neural model was found accurate |
| [63] | EAHE system with a fuzzy logic controller | Authors did not mention any validation process |

## 3. Experimental studies

EAHE systems are used extensively to provide space heating/cooling and air-conditioning in buildings and agricultural greenhouses [20]. Many EAHE experimental studies have been carried out throughout the world. Trombe et al. [113] offered one of the first experimental studies of EAHE for space cooling in the south area of Toulouse (France). Two similar residential buildings were selected for the experiment, one of which was equipped with EAHE for comparative reasons. The main tube characteristics were: (a) PVC material; (b) diameter and thickness of 0.2 m and 5 mm respectively; (c) burial depth of 2.5 m; (d) length of 42 m; and (e) air flowrate equals to 0-45m$^3$/h. Air temperature and relative humidity at the inlet and outlet of the pipe were experimentally measured parameters, with measurements taken every hour. The main result was that the air temperature at the pipe inlet fluctuated between 18 and 36°C for a period of 8 days, while the air temperature variation at the pipe's outlet ranged between 18 and 25°C, underscoring the potential of the system.

In [17] an EAHE system was used to provide air-conditioning in eight rooms of a double-floor guest house in India. A buried pipes system with forced air circulation was used in closed loop mode. The cooling potential of the system was investigated. Multiple heat exchangers were buried parallel to each other, with a distance of 1m between adjacent pipes, at a depth of 2.5m below the ground surface. The exchangers were made of concrete, with a length of 85m, and a diameter of 0.5m. The mean value of air velocity inside the tubes was 6.3m/s. The ambient air temperature, the spatial air-flow velocity at the duct-openings, the spatial indoor air temperature, and the relative humidity of air-conditioned and non-air-conditioned rooms were measured every two hours for a month. The results showed that the ambient air temperature varied between 22.5 and 44.2°C with a relative humidity between 9.4 and 75.8%; the air-conditioned room air temperature fluctuated between 25.3 and 28.4°C, while the non-air-conditioned room air temperature was 2-5°C higher. The indoor relative humidity was 40.7 to 48% for the non-air-conditioned room and 40.8 to 70.3% for the air-conditioned room, providing an acceptable level of comfort.

An investigation of the energy efficiency of several EAHE systems used for space heating and cooling in office buildings in Germany was presented in [64]. Three experimental projects were described, all in Germany. EAHE systems consisted of multiple (from 2 to 26) pipes, with lengths ranging from 67 to 107m, diameters between 0.2 and 0.35m, burial depths between 2 and 4m, air velocity between 1.6 and 5.6m/s, and soil dry-rocky, dry-gravel, and moist-clay soil types. The measurements period included one year for the three experimental projects. The experimental results could be summarized as follows: (a) the annual specific heating energy gain fluctuated between 16.2 and 51.3kWh/m²; and (b) the annual specific cooling energy gain varied between 12.1 and 23.8kWh/m².

Two experimental studies of EAHE systems used for providing heating and cooling in agricultural greenhouses in Delhi (India) were presented in [100] and [66]. EAHE was designed and coupled with the greenhouse for heating during the winter, and cooling during the summer. The climate of the area is representative of tropical climates, with absolute minimum and mean minimum ambient air temperature values during the winter being around 4 to 9°C respectively, and maximum and mean maximum air temperature values for the summer equal to 45 and 39°C, respectively. The used EAHE system consisted of multiple PVC pipes, with a length of 39m and a diameter of 0.06m. The burial depth was 1.2m below the ground surface, covered by bare soil. Pipes were arranged in the ground in a serpentine manner. The ground and indoor air temperature values were recorded, and it was concluded that the indoor air temperature increased by an average of 6 to 7°C during the winter for [100], and up to 4°C for [66]; while in the summer they decreased by 3 to 4°C on the average for [100], and up to 8°C for [66].

A solar greenhouse installed at Izmir (Turkey) was employed for an experimental study of an EAHE underground air tunnel in [114]. The experimental setup consisted of a horizontal U-bend type EAHE system (closed loop), including a galvanized pipe 47m in length and 0.56m in diameter, buried horizontally at a depth of 3m, and a galvanized tube of 15m in length and 0.8m in diameter, coupled with the greenhouse, as illustrated in Figure 2. The soil was a combination of clay, sand, and small rock.

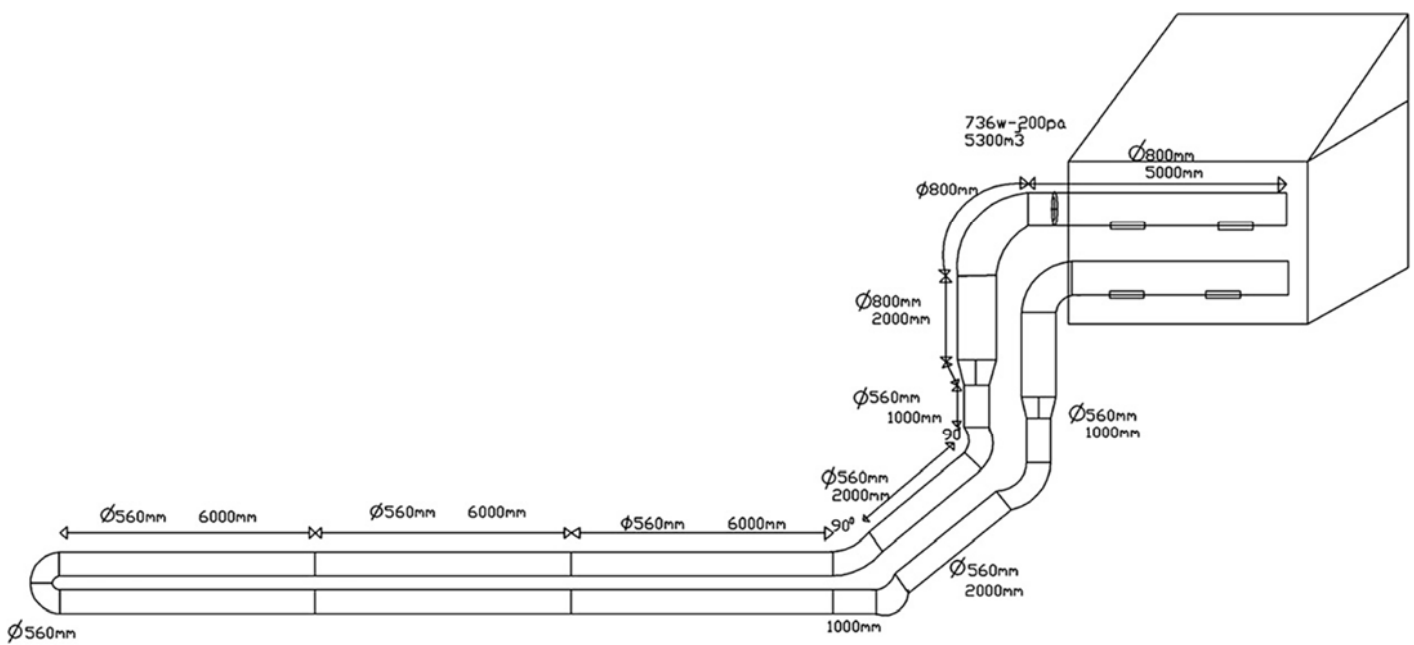

**Figure 2.** Schematic diagram of experimental set up [114]

Those authors measured the temperature and humidity at many locations, including the system inlet and outlet, ambient space, indoor greenhouse, etc. The measured data were used to investigate the exergetic efficiency of the system, based on an energy and exergy analysis presented in the study. The overall efficiency of the exergy system efficiency was found to equal 60.7%.

Vaz et al. [16], [115] developed and validated experimentally a numerical model describing the thermal performance of an EAHE system for space heating/cooling. The experimental study was conducted in a residential building in Brazil; the EAHE system consisted of three buried tubes, A, B, and C. All tubes were made of PVC, with tubes A and B being 0.11m in diameter and buried at a depth of 2m below the ground surface, and tube C being 0.10m in diameter and buried at a depth of 0.5m. Experimental results showed that for a burial depth of 2m the temperature of the air flowing inside the tube had the potential to rise more than 8°C or drop by about 4°C. An extension of the experimental investigation described in [16], intended to assess the heating/cooling potential of an EAHE system in the same residential building in Brazil, was performed in

[115]. As in [16], the system consisted of three ducts, A, B, and C, all buried at different depths (1.60, 0.60, and 0.50m correspondingly). Those authors found that the heating/cooling potential of the system was significant and could remarkably improve the thermal comfort conditions inside the building. Moreover, the results showed that the best periods for using the system were the months of February for cooling and May for heating.

An experimental ventilation system for space heating/cooling was used to validate a relevant numerical model [22]. The system consisted of six tubes, buried horizontally at a depth of 1.5 to 3m below the ground surface, 50m in length and 0.4m in diameter. Experimental results may be summarized as follows: (a) the system could provide 62% of the heating requirements for the month of March, with a coefficient of performance (COP) of 3.2; (b) the system could provide 86% of the cooling load for the month of July, with a COP of 3.53.

Mongkon et al. [116] carried out an experimental investigation of a horizontal EAHE system to provide cooling in an agricultural greenhouse in Thailand. The EAHE system was buried at a depth of 1m below a ground surface covered with a short lawn. The system consisted of an iron tube, placed in six serpentine shape rows, as illustrated in Figure 3. Inlet and outlet tubes were made of PVC, and had a diameter of 0.08m. Experimental results showed that the system had a significant cooling efficiency for tropical climates.

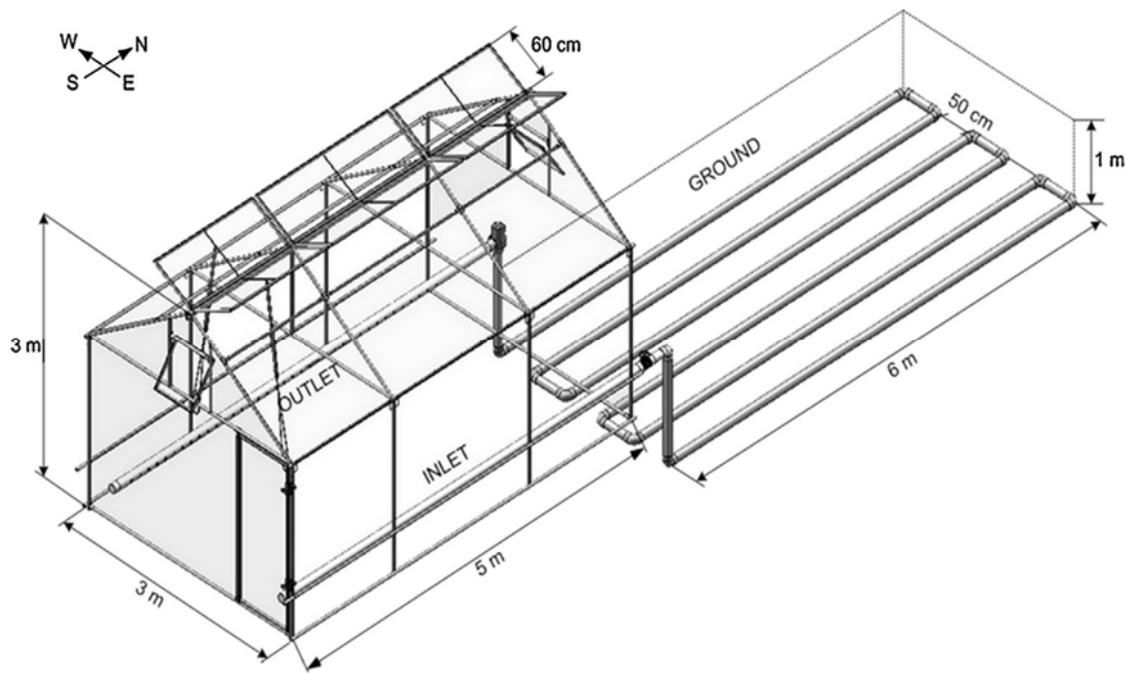

**Figure 3.** Schematic diagram of experimental set up [116]

Chiesa et al. [67] presented a case study of monitoring the performance of an extensive EAHE system located in a school building in Imola (Italy), providing space heating and cooling, as shown in Figure 4. The EAHE system consisted of three sets of pipes buried in three fields. Two sets had 12 pipes each, while the third set had only eight pipes. The pipes were 70m in length and 0.25m in diameter. The average burial depth below the ground surface was 2.61m while the distance between adjacent pipes in the same field was 1.10m. The average air velocity inside the tubes was 2m/s. Those authors presented results of 12-month monitoring for all three fields of the EAHE system, which was proven remarkably effective for cooling and pre-heating modes.

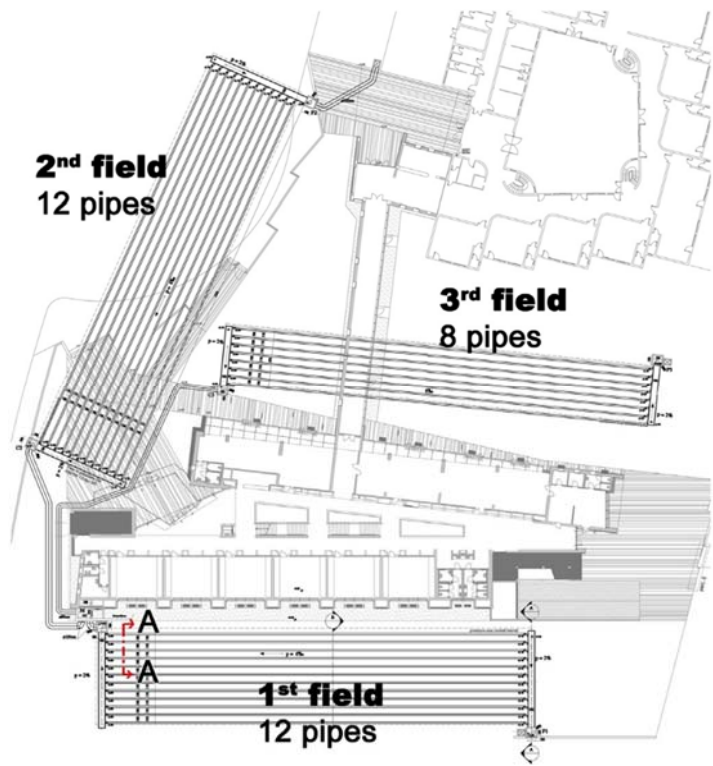

**Figure 4.** Plan of installed EAHE system [67]

Serageldin et al. [41] carried out an experimental study to validate their numerical model based on computational fluid dynamics. A 5.5m long horizontal EAHE PVC pipe with a diameter of 0.05m was buried in a serpentine shape at a depth of 2m. Air circulated inside the tube at a velocity of 1 to 3.9m/s while a loamy sand soil was selected. Experiments were performed in winter and summer, and the results showed a significant heating/cooling potential, which depended strongly on the design parameters of the system.

An earth-to-air heat exchanger system was used experimentally for cooling a residential building in Marrakesh (Morocco) [68]. The system consisted of three parallel U-shaped tubes, 77.7m in length and 0.15m in diameter, buried at a depth of 2.2 to 3.5m below the ground surface. Two of the pipes were coupled with the first floor of the building, while the third was connected with the second floor. The results showed remarkable system efficiency, offering an almost constant pipe outlet air temperature of 25°C, when the ambient air temperature was over 40°C.

An experimental investigation carried out to assess the cooling potential of an EAHE system for hot and arid climates, was presented in [117]. The experimental setup, depicted in Figure 5, consisted of a pipe of 1.5 m in length and 0.15m in diameter buried in soil contained in a galvanized steel drum. The system was connected with a blower and an air heater to circulate warm air inside the tube. The results showed that the system could be remarkably efficient, offering an outlet air temperature reduction of up to 24°C.

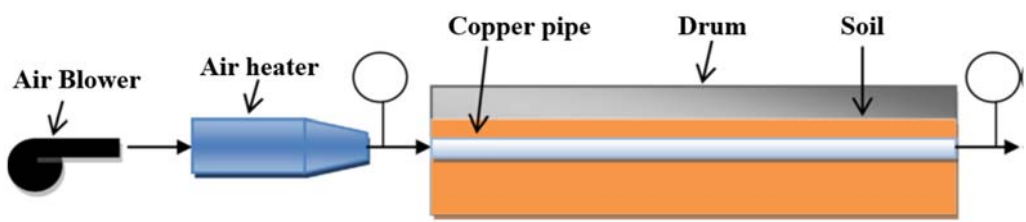

**Figure 5.** Schematic diagram of experimental set up [117]

Yusof et al. [118] employed an innovative laboratory simulator to assess the energy efficiency and the potential of an EAHE. The simulator acted as real ground surrounding the pipes, and allowed operation in steady state, eliminating the accumulation term in the energy balance equation. The EAHE laboratory simulator included an 8.7m long and 101.8mm inside diameter PVC pipe. The results showed that efficiency ranged up to 88%, while the air temperature reduction reached around 10°C, thus achieving a temperature reduction of 27.5% compared with the air temperature at the pipe inlet. Liu et al. [30] carried out an experimental study to validate their transient numerical model for predicting the thermal performance of a vertical EAHE system. Experiment was performed in Changsha (China) which has a subtropical climate (hot summer and cold winter). The system consisted of a U-shaped pipe (U-tube), made of stainless steel, which was 0.219m in diameter, buried into a 16.5m deep hole with a 1m diameter, filled with soil. Measurements were collected during both winter and summer. The authors were in favor of the Vertical EAHE (VEAHE), and experimental results were used to validate their model, so as to further study the parametric sensitivity of the EAHE system. As such, they studied the effect of tube material, tube insulation, air-flow, and soil types. Their results indicated that optimum choices depend on specific objectives. In [69], an extension of experimental and theoretical research on VEAHE was presented and discussed. Those authors argues that U-shaped tubes presented two main advantages: (a) a significant reduction in their requirement of land use; and (b) an improvement of system efficiency due to increased burial depth, as shown in Figure 6. The location of the experiment was again Changsha (China), and the experimental set up consisted of the U-tube buried in the hole at a depth of 16.5m. Experimental results showed that the system can be very efficient in both summer and winter. The air temperature at the pipe outlet fluctuated between 22.4 and 24.4°C in the summer, and between 16 and 18°C in the winter. Moreover, VEAHE could exploit the quality of deep ground as a natural heat sink for the cooling period, as well as a natural heat source for the heating period of the year.

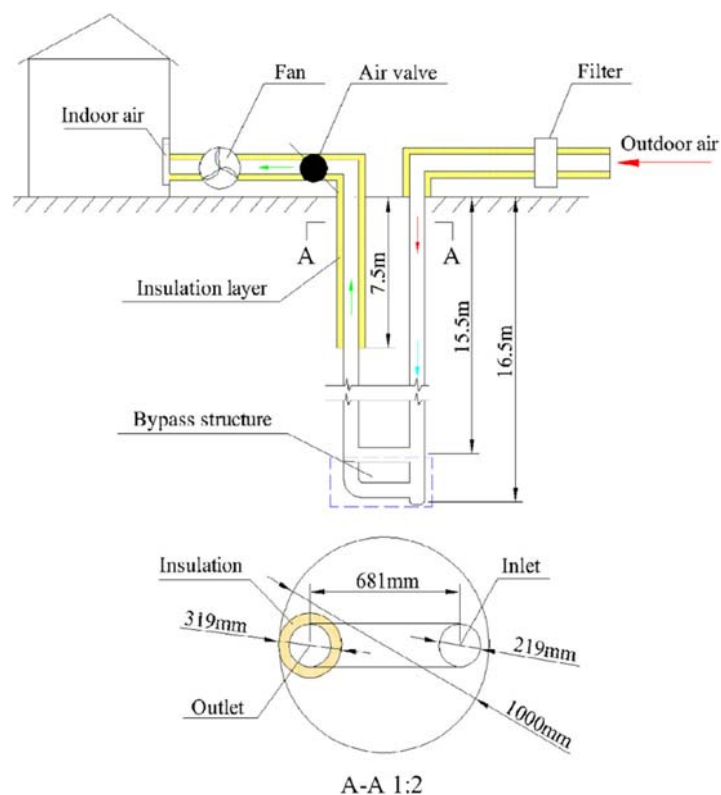

**Figure 6.** Schematic diagram of experimental set up [69]

In [119] a full-scale experimental study was carried out to assess the cooling potential of an EAHE system in a city in southwest China, characterized by hot and humid climate. The experimental setup consisted of four buried PVC pipes of different design parameters (diameter of 0.16, 0.11, 0.11 and 0.075m correspondingly). All pipes had a vertical inlet, a horizontal section where the heat transfer process was conducted, and a vertical outlet. The burial depth for three pipes was 3m below the ground surface, and 2m for the fourth one. Adjacent pipes buried at 3m were 0.7m apart. Experimental results showed that the EAHE system presented a remarkable cooling capacity and could significantly cool and dehumidify the ambient air. The outlet air temperature and humidity presented a considerable stability, although the inlet air temperature and humidity fluctuated between 21.5 and 41.2°C, and from 11.2 to 20.5g/kg for an EAHE of a substantial length. The achieved maximum air temperature reduction for the buried pipe of 0.075m in diameter buried at a depth of 3m, was 22.1°C, while for moisture content the reduction was 7.41g/kg. Sensible heat transfer capacity varied in the range of 60 to 83%, while the cooling capacity for latent heat transfer fluctuated between 17 and 40%.

EAHE systems are multiparametric. The efficiency of the studied systems directly depended upon the environmental conditions and the engineering design, and their conclusions were relevant only for the specific conditions and design. Nevertheless, the aforementioned studies, shown in Table 2, were conducted in different places around the globe with different thermophysical properties and engineering choices. In spite of the fact that conclusions from individual research works cannot be extrapolated, there is some common ground and their findings converge to the following three facts: (1) they demonstrated that buried pipes are mostly affecting summer indoor temperatures (tropical cooling); (2) all system efficiency values fluctuated around 60%; and (3) the outlet air temperature did not guarantee a rise of the indoor temperature in the winter, but it appeared to efficiently reduce summer indoor temperatures. Thus, the above studies seem to suggest that EAHE are more efficient as a means of cooling rather than a method of heating.

Irrespective of the efficiency of those systems, they were not optimized on the needs of indoor air quality. Most research was conducted on empirically designed systems. Although the reported research has provided encouraging results which could find direct application without any further refinement in order to reduce energy requirements in buildings, the authors of this work consider that the need for research which probes into the science behind such applications is imperative. As such, it is recommended that research be conducted in laboratory simulators, where the effect of each parameter is evaluated in isolation or in groups. At the same time, fundamental experimental research on temperature profiles as a function of the distance from the earth surface, local climate, weather conditions, soil type, earth vegetation, and annual rainfall can be useful in selecting the most appropriate engineering design. Finally, a life cycle assessment is also necessary, e.g. the energy employed for the fans, digging the ground, manufacturing the tubes, etc., should not exceed the heating/cooling energy savings.

Table 2 offers a summary of the main characteristics and results of representative implementations of EAHE systems in various types of buildings, greenhouses and research laboratories.

**Table 2.** Summary of main characteristics and results of experimental studies of EAHE systems

| | EXPERIMENTAL STUDIES | | | |
|---|---|---|---|---|
| **Reference** | **EAHE configuration parameters** | **Experimental location and Building type coupled with EAHE system** | **Operation mode** | **Main Results** |
| [113] | Material: PVC; length: 42.0m; diameter: 0.2m; depth: 2.5m. The system was buried under the building garden | Residential building in the south of France | Space cooling | a) Air temperature variation at the pipe's inlet: 18-36°C<br>b) Air temperature variation at the pipe's outlet: 18-25°C |
| [17] | Multiple parallel tubes; material: concrete; length: 85m; diameter: 0.5m; air velocity: 6.2m/s; tube spacing: 1m | Double floor guest house in India | Space cooling | a) Ambient temperature variation: 22.5-44.2°C; ambient relative humidity variation: 9.4-75.8%<br>b) Room indoor air temperature variation: 25.3-28.4°C; indoor relative humidity variation: 40.8-70.3% |
| [64] | Multiple pipes; lengths: 67-107m; diameters: 0.2-0.35m; burial depths: 2-4m; air velocities: 1.6-5.6m/s; soil types: dry-rocky, dry-gravel, and moist clay. | Office building in Germany | Space heating and cooling | a) The specific heating energy gain fluctuated between 16.2 and 51.3 kWh/m$^2$ per annum<br>b) The specific cooling energy gain fluctuated between 12.1 and 23.8 kWh/m$^2$ per annum |
| [65] | Multiple pipes; length: 39m; diameter: 0.06; depth: 1.2m; pipes were arranged in a serpentine fashion | Agricultural greenhouse in Delhi (India) | Heating and cooling | Indoor air temperature values increased by an average of 6-7°C during the winter and decreased by an average 3-4°C in the summer |
| [66] | Multiple pipes; length: 39m; diameter: 0.06m; depth: 1.2m; pipes were arranged in a serpentine fashion | Agricultural greenhouse in Delhi (India) | Heating and cooling | Indoor air temperature increased by up to 4°C during the winter while in the summer the use of buried pipes resulted in a temperature decrease reaching 8°C |
| [114] | Horizontal U-bend type EAHE system (closed loop); length: 47m; diameter: 0.56m; depth: 3m; soil was a combination of clay, sand and gravel | Experimental solar greenhouse at Izmir (Turkey) | Heating and cooling | Overall exergy system efficiency found to be 60.7% |
| [16], [115] | Multiple pipes with different design parameters; material: PVC | Residential building in Brazil | Heating and cooling | Heating/cooling potential of the system was significant and could improve the thermal |

| | | | | comfort conditions inside the building remarkably. The best period for using the system was the month of February for cooling, and the month of May for heating |
|---|---|---|---|---|
| [22] | Six horizontal tubes; length: 50m; diameter: 0.4m; depth: 1.5-3m | Research Laboratory, University of Nottingham, Ningbo, China | Heating and cooling | Heating mode: system would provide 62% of the heating load for the month of March with a COP of 3.2. Cooling mode: system could provide 86% of the cooling load for the month of July with a COP of 3.53 |
| [116] | Horizontal tube, made of iron, placed in 6 serpentine shaped rows. Inlet and outlet tubes were made of PVC, 0.08m in diameter. The system was buried under short-lawn ground surface. | Agricultural greenhouse in Thailand (tropical climate) | Cooling | System presented significant efficiency for cooling in tropical climates |
| [67] | Three fields of buried pipes with different number of pipes (12, 12 and 8), 70m in length and 0.25m in diameter; depth: 2.61m; air velocity: 2m/s; distance between adjacent pipes: 1.10m | School building in Imola, Italy | Heating and cooling | System proved remarkably effective for cooling and pre-heating modes |
| [41] | Horizontal EAHE in serpentine shape; material: PVC; inner diameter: 0.05m; length: 5.5m; air velocity: 1-3.9m/s; soil: loamy sand | Energy Resources Engineering Department Laboratory, University of Science and Technology, Egypt | Heating and cooling | Significant heating/cooling potential strongly depended on system design parameters |
| [68] | Three parallel U-shaped tubes; length: 77.7m; diameter: 0.15m; depth: 2.2-3.5m | Residential building in Marrakesh (Morocco) | Cooling | System presented remarkable efficiency offering almost constant pipe outlet air temperature values of 25°C when the ambient air temperature was higher than 40°C |
| [117] | One pipe; length: 1.5m; diameter: 0.15m; system was connected with a blower and an air heater to circulate warm air inside the tube | Application for hot, dry, and arid climates | Cooling | System could be remarkably efficient, offering an outlet temperature reduction up to 24°C |
| [118] | EAHE system laboratory simulator; pipe made of PVC; length: 8.7m; inside diameter: 101.8mm | Implementation in tropical climates | Cooling | System efficiency ranged up to 88% and the air temperature reduction reached 9.62°C, achieving a temperature reduction of 27.5% |

| | | | | compared with the air temperature at the pipe inlet. |
|---|---|---|---|---|
| [30] | U-shaped pipe (U-tube) made of stainless steel; diameter: 219mm, buried into a deep hole of 16.5m depth and 1m diameter, filled with soil | Experiment carried out in Changsha (China), a city of sub-tropical climate (hot summers and cold winters) | Heating and cooling | Experimental validation of numerical model |
| [69] | U-shaped tubes (U-tubes), as in [30] | Experiment carried out in Changsha (China), a city of sub-tropical climate (hot summers and cold winters) | Heating and cooling | Air temperature at the pipe outlet fluctuated between 22.4°C and 24.4°C in the summer and 16.0°C to 18°C in the winter |
| [119] | Four pipes made of PVC with different design parameters; depth: 3m (three pipes), and 2m (one pipe);. distance between adjacent pipes: 0.7m | City in south-west China characterized by hot and humid climate | Cooling | EAHE system presented remarkable cooling capacity and was able to cool and dehumidify the ambient air significantly. System cooling capacity as regards sensible heat transfer varied between 60.5% and 82.82%, while the cooling capacity for latent heat transfer fluctuated between 17.18% and 39.5%. |

## 4. Parametric studies

The research literature on EAHE systems includes many parametric studies expressed in the form of sensitivity analysis and investigation of the impact of the main system design and environmental parameters on the space cooling/heating efficiency. The thermal behavior of an EAHE is strongly affected by many parameters of different type and quality, which were classified into three main categories: (a) buried pipes system design parameters; (b) soil types described by the thermophysical characteristics of the ground; and (c) environmental parameters, mostly influencing the ground surface temperature distribution. Parameters in the first category include: pipe length; pipe radius; air velocity inside the tube; burial depth; number of pipes; and pipe material. Those in the second category include: soil moisture content; soil density; soil specific heat capacity; and soil thermal diffusivity. Finally, those in the third category include: short and long wave radiations; convective heat flux between air and ground surface; latent heat flux due to evaporation/condensation; wind speed; vegetation; and different ground covers. Most researchers have investigated the effect of parameters in the first category (system design) on the cooling/heating capacity of EAHE systems, although there are also studies including parameters of the other two categories. Table 3 lists studies addressing the effect of crucial parameters on the thermal efficiency of an EAHE system.

**Table 3.** Parametric studies of the effect of crucial parameters on the thermal efficiency of an EAHE system

| PARAMETRIC STUDIES | | |
|---|---|---|
| **Reference** | **Category of Parameters** | **Investigated Parameters** |
| [70] | (a), (c) | Pipe length, radius, air mass flow rate, ground surface cover |
| [120] | (a) | Pipe length, radius, air velocity, burial depth |
| [121] | (a) | Pipe length, radius, air velocity, burial depth |
| [43] | (a) | Pipe length, radius, air velocity, burial depth, space between pipes |
| [71] | (c), (a) | Ground surface cover, pipe length, radius, burial depth, air velocity |
| [122] | (c), (a) | Ground surface cover, pipe length, radius, burial depth, air velocity |
| [123] | (a) | Pipes length, pipes radius, burial depth, and air velocity inside the pipes |
| [124] | (a) | Pipes length, pipes radius, burial depth, and air velocity inside the pipes |
| [109] | (a) | pipe diameter, air velocity inside the tube |
| [60] | (a), (b) | Pipe length, diameter, burial depth, air velocity, and soil types |
| [47] | (a), (c) | Pipe diameter, burial depth, pipe material, and ground surface cover |
| [57] | (a) | Pipe radius, length, air flow rate, and burial depth |
| [125] | (a) | Burial depth, pipe length, air velocity inside the tubes, and pipe material |
| [126] | (a) | Pipe material, length, burial depth, air velocity inside the pipe, and soil types |
| [36] | (a) | Burial depth, form factor sigma, and Reynolds number |

| [58] | (a) | Pipe length, burial depth, and air velocity |
| [59] | (a) | Burial depth, pipe radius, length and ventilation flow rate in the pipe burial depth, pipe radius, pipe length and ventilation flow rate in the tube |
| [108] | (a) | Air velocity inside the tube, pipe length, diameter, burial depth, and pipe material |
| [28] | (a) | Pipe length |
| [72] | (a) | Pipe material, and burial depth |
| [119] | (a) | Burial depth, air velocity inside the tubes, and pipe length |
| [73] | (b) | Soil types |
| [74] | (c) | Ground surface cover |

Sodha et al. [70] investigated the effect of pipe length, radius, and air mass flow rate inside the tube, (first category); as well as earth surface coverage (third category), on the EAHE system efficiency for space cooling for hot and dry climates in India. Those authors presented the most effective solution as regards length, radius, mass flow rate, and number of pipes for the cooling period of the year. As regards different ground surface covers, sunlit, shaded, wetted, and wet shaded were considered, with wet shaded found to constitute the most effective solution.

Four main system design parameters, pipe length, pipe radius, air velocity inside the pipe, and burial depth have been regarded as key-variables for extensive sensitivity analyses for a single pipe and for space cooling/heating in [120] and [121] respectively, and in [43] for multiple pipes and space cooling. The system cooling/heating capacity was shown to be strongly related to the previous design parameters. For the single pipe, an increase of pipe length and burial depth resulted in an increase of system efficiency, while an increase of pipe radius and air velocity resulted in a reduction of system capacity. For multiple pipes, the distance between adjacent pipes was also considered, and it was found that an increase of the spacing between the pipes led to a slight reduction of air temperature at the pipe exit, when the system operated for space cooling.

Different ground surface covers and an environmental parameter (of the third category) were the main factors influencing the EAHE system efficiency, investigated and presented in [71] for space cooling and [122] for space heating. Two different ground surface boundary conditions were considered: (a) bare soil, and (b) short-grass soil. The cooling potential of the system was investigated and assessed in both cooling and heating operation modes. An extensive sensitivity analysis including the main design parameters of the system (pipe length, radius, burial depth, and air velocity inside the pipe) was presented. Research showed that short grass soil, as ground surface coverage, resulted in a significant improvement of the cooling capacity of the system, while bare soil was found to increase the system's heating capacity.

Multiple pipes were used for cooling and heating a glass-covered 1000m² agricultural greenhouse in Athens (Greece) in [123] and [124] respectively. Sensitivity analyses were performed for cooling and heating operation modes and for the system's design parameters (length and radius of pipes, burial depth, and air velocity inside the pipes). It was shown that for cooling, the greenhouse indoor air temperature decreased with the increase of pipe length and burial depth, while the indoor air temperature increased with the decrease of pipe radius and air velocity inside the tubes. For the heating mode, the system's heating capacity increased with increasing pipe length and burial depth, and decreased with increasing pipe radius and air velocity inside the tubes.

A parametric space cooling investigation was presented in [109], concerning the influence of two system design parameters: pipe diameter, and air velocity inside the tube. It was found that both parameters had a remarkable effect on the cooling capacity of the system, thus an increase of pipe diameter and air velocity resulted in a significant increase in the air temperature at the pipe exit.

Deglin et al. [60] developed a detailed analysis of the effect of the following system parameters: length, diameter, burial depth, and air velocity inside the tube (first category); and soil type (second category). Two different types of soil were selected: dry sand, and saturated silt; it was found that the most efficient soils were those with higher saturation, thus higher conductivity. As regards the design parameters of the system, the effect of length was significant as greater length provided better results at the pipe exit, while infinite length provides 100% efficiency, with cost a limiting factor. For diameter and air velocity, small pipes were more efficient, while decreasing air velocity resulted in improved efficiency. As regarded burial depth, an increase of depth decreased the influence of the fluctuations of the ground surface temperature, again with cost being a limiting factor.

In [47], a sensitivity analysis was developed regarding the following parameters: pipe diameter, burial depth, and pipe material (first category); and ground surface cover (second category). Heating and cooling efficiency was shown to increase with increasing burial depth, and decreasing with increasing diameter. Plastic, aluminum, concrete, and roughcast were used as pipe materials, and their effect on system efficiency was found to be insignificant. Finally, as regards ground surface coverage, those authors selected a short grass-covered, and an asphalt-covered soil. For a heating operation mode, bare soil was found to increase the efficiency of the system because of its high absorptance.

Lee et al. [57] carried out a parametric investigation of the effect of the following parameters of the first category on the thermal capacity of the system for cooling operation mode: pipe radius, length, air flow rate, and burial depth. Those authors concluded that, for cooling mode, lower air temperature values at the tube exit were achieved with longer pipes, bigger depths, smaller diameters, and lower air velocities inside the tubes. Evidently, cost also played an important role.

The effect of parameters of the first category, such as burial depth, pipe length, air velocity inside tubes, and pipe material on the thermal efficiency of an EAHE system was investigated and analyzed in [125]. Those authors found that all system parameters could affect the cooling capacity of the system more than the heating one. A galvanized pipe and a PVC were used, and it was found that outlet air temperature differences were insignificant for both cooling and heating operation modes. An extensive sensitivity analysis was developed and presented in [126], aimed at investigating and analyzing the EAHE heating and cooling potential for different Italian climates, studying the following parameters of the first category: tube material, tube length, burial depth, and air velocity inside the pipe. It was shown that the effect of pipe material, (PVC, metal, or concrete) on the EAHE performance was negligible. As regards the system design parameters, the preferred choices were 50m of pipe length, at a burial depth of 3m, and an air velocity inside the tube equal to 8m/s.

Sehli et al. [36] carried out a parametric study for cooling, considering burial depth, form factor $\delta$ (equal to pipe length over diameter), and the Reynolds number representing the air flow inside the tube. It was shown that an increase of burial depth and form factor resulted in lower air temperature values at the outlet, while an increase of Reynolds number resulted in an increase of the air temperature at the outlet.

A sensitivity analysis was performed in [58] to study the effect of pipe length, burial depth, and air velocity inside the tube, on the thermal performance of an EAHE system operated in a closed loop mode, and used for space cooling in hot and humid climates. Pipe length, burial depth, and air velocity were found to play an important role on the system efficiency, as the cooling capacity of EAHE increased with longer tubes, increased depth, and smaller velocities. Those authors proposed to take these parameters into account simultaneously in order to optimize the performance of the system.

In an extensive parametric study, Yang et al. [59] analyzed the influence of the following (first category) system design parameters on the air temperature fluctuation at the EAHE exit: burial depth, pipe radius, pipe length, and ventilation flow rate in the tube. It was confirmed that both pipe length and burial depth are are the most significant systems' parameters. In particular, the increase of both pipe length and burial depth result in a remarkable improvement of the system's cooling capacity. An increase of the pipe radius and the air flow rate resulted in a slight increase of outlet air temperature.

Ahmed et al. [108] offered an extensive sensitivity analysis related to the thermal performance of a horizontal EAHE system for space cooling. The considered (first category) parameters were: air velocity inside the tube; pipe length, diameter, and material; and burial depth. As regards air velocity, four values were considered: 0.41, 1.0, 1.5, and 2.0m/s; it was found that 1.5m/s offered the best cooling capacity. Pipe lengths of 7.5, 15, 30, and 60m were considered, and the 60m gave the optimum performance with velocity equal to 1.5m/s. The smallest pipe diameter offered the most improved performance. As regards pipe material, PVC, polyethylene, concrete, and clay were used, and it was found that the material with the highest conductivity gave the most improved cooling performance. Finally, burial depths of 0.6, 2.0, 4.0, and 8m were used in simulations, and cooling effect was found to increase with depth.

In an EAHE study, Belatrache et al. [28] investigated the effect of pipe length on the air temperature difference between inlet and outlet air during the summer period. The air temperature difference between the inlet and outlet of a pipe was found to increase significantly with length up to about 25m. For bigger lengths, the temperature difference became insignificant. In [72], a study of the influence of two (first category) parameters, pipe material and burial depth, on the thermal efficiency of an EAHE system was undertaken. Two pipe materials were considered, steel and PVC, and it was shown that the material selection did not offer a significant effect on system efficiency. For burial depth, those authors concluded that a depth of 100cm would be an optimal choice for an improved system performance.

Wei et al. [119] carried out an extensive experimental investigation of the cooling capacity of an EAHE system for hot and humid climates. Those authors also carried out a sensitivity analysis using the following (first category) system parameters: burial depth, air velocity inside the tubes, and pipe length. An increase of burial depth resulted in improving drastically the air temperature at the outlet, and the cooling capacity of the system. Decreasing the pipe diameter resulted in minimizing the outlet air temperature and humidity, but it did not improve the cooling capacity of the system. An increase of air velocity resulted in an increase of outlet temperature for the same length, while the air temperature at the pipe outlet decreased exponentially with increasing pipe length.

An investigation of the influence of soil thermal conductivity on the thermal behavior of an EAHE system for space cooling was done in [73]. Soil thermal conductivities equal to 0.52, 2, and 4 $WK^{-1}m^{-1}$, resulted in a maximum air temperature drop of 15.6, 17.0 and 17.3K. The type of soil coating, which was tested in [74] to calculate its effect on the thermal performance of EAHE for cooling and heating, was found to play a significant role on the thermal performance of a system. Three types of soil coating, sand, sand-bent (a mix of sand and bentonite), and in-situ earth, were combined with two types of moisture (minimum and maximum). The system thermal performance presented a difference of up to 15.9% for the same soil coating with the two different moisture contents, and a difference of up to 17.4% for minimum moisture with the different soil coating types. For the maximum moisture content, the soil coating type resulted in an impact of only 2% on the thermal performance. A summary of the aforementioned parametric studies is shown in Table 4.

**Table 4.** Summary of the results of parametric studies

| PARAMETRIC STUDIES RESULTS | | |
|---|---|---|
| **System's Parameter** | **Reference** | **Results** |
| Pipe length | [28], [36], [43], [57]–[60], [70], [71], [108], [119]–[126] | An increase resulted in a significant improvement of system efficiency |
| Pipe radius/diameter | [36], [43], [47], [57], [59], [60], [70], [71], [108], [109], [120]–[124] | An increase resulted in a reduction of system efficiency |
| Air velocity inside the tube/air mass rate | [43], [57]–[60], [70], [71], [108], [109], [119]–[126] | An increase resulted in a reduction of system efficiency |
| Burial depth | [36], [43], [47], [57]–[60], [71], [72], [108], [119]–[126] | An increase resulted in a significant improvement of system efficiency |
| Pipe material | [47], [72], [108], [125], [126] | No significant impact on the system efficiency |
| Space between adjacent pipes | [43] | An increase resulted in an improvement of system efficiency |
| Ground surface cover | [47], [71], [74], [122] | Short-grass covered soil improved the system's cooling capacity while bare soil improved the heating one. Soil surface cover played an important role in the EAHE system thermal performance. |
| Soil types | [60], [73], [126] | Higher conductivity soils were the most efficient |

## 5. Hybrid EAHE-renewable systems

In all EAHE systems, a small amount of electricity is necessary to transfer both thermal and cooling load to specific locations within buildings and structures. This is also necessary to meet the needs of air conditioning. The required electrical energy varies according to the type of size applications. In addition, the possibility of coupling EAHE systems with energy systems producing heat is a particularly effective solution to reach maximum heat production with maximum efficiency.

In the previous sections of this work, a number of the discussed types of EHAE systems referred to their autonomous operation, using conventional power sources. In this section, recent research papers that confirm the fact that the effectiveness and utility of an EAHE system may be increased by coupling it with active Renewable Energy Systems (RES), are discussed. This way, a hybrid system is defined, which consists of a specific type of EAHE in combination with several renewable energy (RE) technologies so that the requirements of thermal loads may be addressed, something that a standalone EAHE would be unable to do.

In a study published in 2019, Agrawal et al. [2] discussed the state-of-the-art for applications, technology integration, and the latest research trends on earth-air-heat exchanger systems. Among the studied systems, there were recent works of importance, including hybrid EAHE systems coupled with solar energy systems. Elminshawy et al. [80] used EAHE system for cooling solar (photovoltaic, PV) panels. In particular, they used an EHAE system that cooled a hybrid photovoltaic thermal (PVT) system with air heat exchanger on its rear side, as depicted in Figure 7. That study revealed that, by passing the EAHE cooled air to a solar PV, the temperature of the solar PV module decreased by almost 24% and the electrical conversion efficiency and output power of the PV module increased by 23% and 19%, respectively.

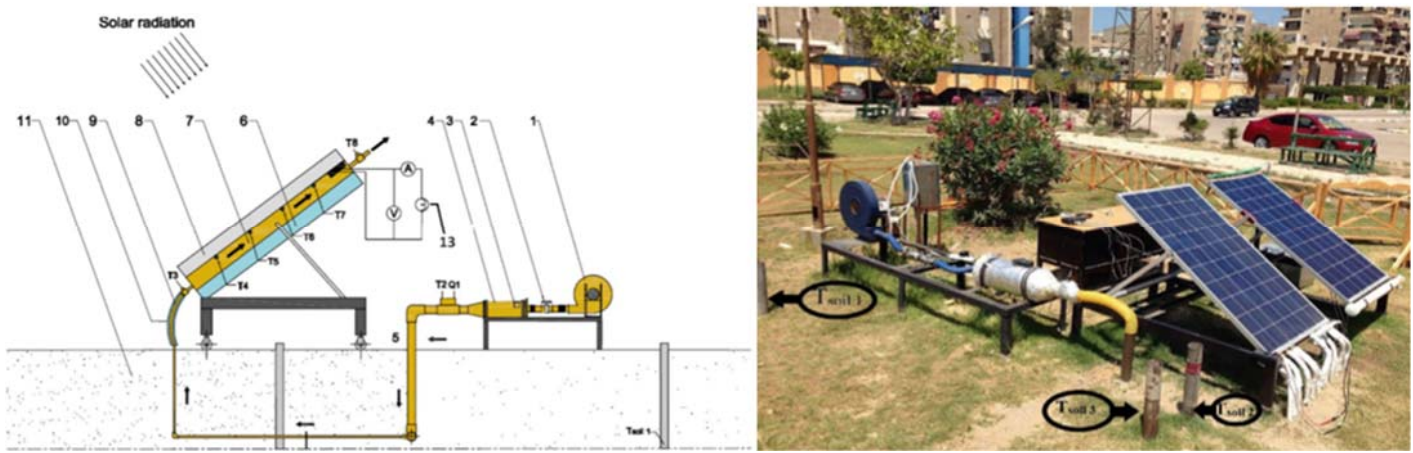

**Figure 7.** Schematic diagram and experimental setup of hybrid EAHE coupled with PVT collector [80]

Following the concept of the operational combination of EAHE and PVT systems, Jakhar et al. [81] studied the effectiveness of an earth-water heat exchanger system in cooling the PV panels. In that study, the maximum PV panel temperature was almost 75°C without cooling, while it decreased remarkably down to almost 46°C with EHAE cooling, for low values of water flow rate.

Most research workers rarely touch upon the improvement of performance of air-cooled PVT systems employing an active geothermal air cooling using a buried EAHE in closed cycle with the (already existing) standalone PV. Yildiz et al. [82] investigates a hybrid EAHE system coupled with PV cells, focusing on the overall performance improvement. That work analyzed the PV assisted closed-loop EAHE exchanger for cooling a greenhouse. The results indicated that 31% of electric energy could be obtained from solar PV cells and the rest 69% of obtained from conventional resources, resulting in total energy savings of almost 30%.

Zapałowicz and Opiela [83] proposed a hybrid air-conditioning system for a building. An EAHE and a PV unit gave a more sophisticated solution. The ambient air was first cooled through the EAHE system, and some of this cooled air flew through a channel formed by the rear side of the PV modules and the wall of the building, reducing the temperature of the PV modules and improving their efficiency. At the same time, the building wall also cooled down and reduced the heat gain into the building. The remaining cooled air was directed to the air conditioner to lower the air conditioning load, as shown in Figure 8.

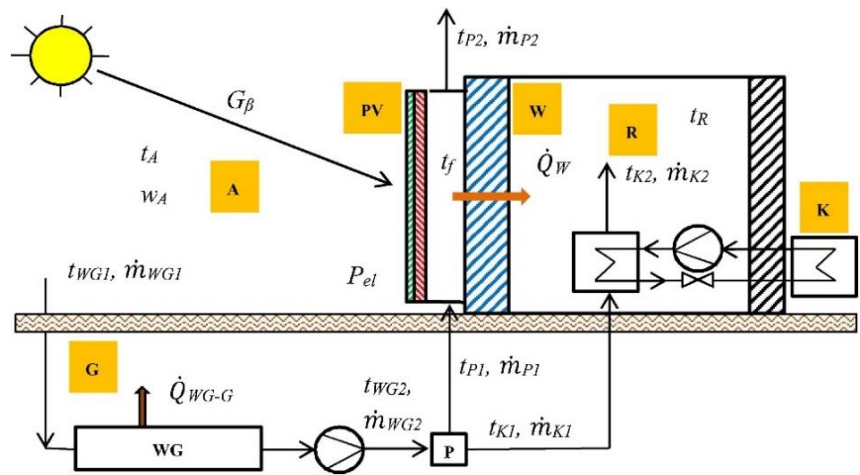

**Figure 8.** Schematic diagram of proposed hybrid EAHE installation [83]

Hachchadi et al. [84] presented a similar performance analysis, which described a PV thermal air collector combined with water-to-air heat exchanger for renewed air conditioning in buildings. The operational of the hybrid system was improved, minimizing the required area of the heat exchanger by about 33%.

On the use of PV or hybrid PVT units in EAHE systems, recent efforts at optimization have been conducted, looking for the optimum values of all parameters. Li et al. [85] accomplished a multi-objective energy and exergy optimization analysis for different configurations of hybrid earth-air heat exchanger coupled with a Building Integrated Photovoltaic/thermal system (BIPVT). The multi-criteria decision making procedure was performed to maximize the yearly total energy and exergy outputs of different configurations. The length, depth, and width of the channel located under the PV modules, air mass flow rate, and the length and diameter of the EAHE system were the decision variables.

Active solar thermal combined with EAHE systems, have proven to be the most promising solution for the effective operation of several types of heat exchangers. Many researchers have examined this coupling over the last decade. Wang et al. [86] studied experimentally an underground thermal storage in a solar-ground-coupled heat pump system for residential buildings. That study implied that the performance of underground thermal storage of a hybrid EAHE is strongly dependent on the intensity of solar radiation, and the matching between the water tank volume and the area of solar collectors. The overall efficiency of the hybrid system improved over 20%, while the efficiency of the absorbed solar energy exceeded 70%. It is obvious that such hybrid EAHE systems operate more effectively than conventional ones.

A multi-objective optimization of a seasonal solar thermal energy storage system combined with an EAHE exchanger and a solar collector field was researched in the work of Benzaana et al. [87], where the produced thermal energy goes to net zero energy buildings (NZEBs). Figure 9 depicts the proposed installation conceptually, considering the basic parameters. Experimentally validated numerical results showed that the combination of the hybrid EAHE system satisfied all energy needs, covering 45% of the total thermal energy all year long.

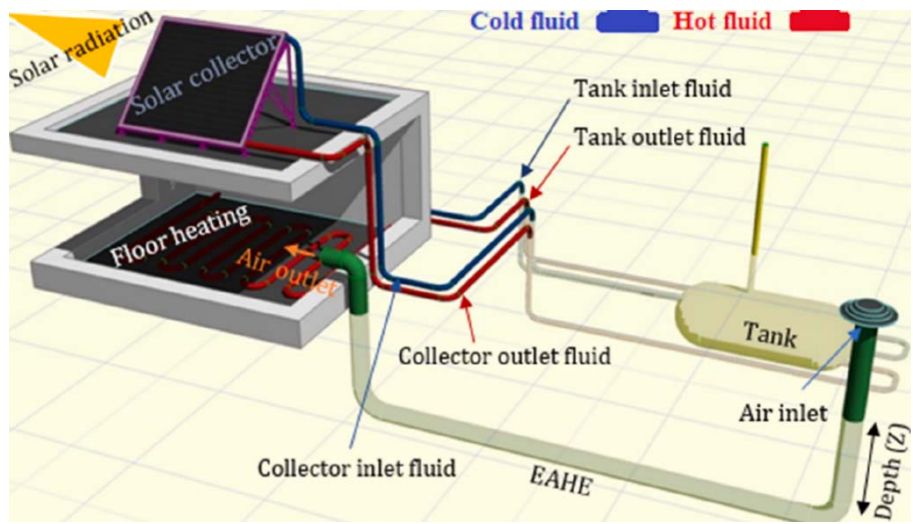

**Figure 9.** EAHE system coupled with underground tank and solar collector for heating [87]

A similar approach was attempted earlier by Attar et al. [88], who used a combination of parametric and numerical study of a solar system for heating a greenhouse equipped with a buried exchanger. In that study, a TRNSYS simulation project was designed, following a determination process for the experimental input parameters in the TRNSYS environment. The results were experimentally validated and showed that the volume of the storage tank in connection with the absorbing area of the solar collector field are the most significant parameters. In particular, the lower flow rate in the heat exchanger and the minimization of the temperature stratification of the storage tank could increase the air temperature inside the greenhouse by 5°C.

Passive cooling of buildings through geothermal systems has been one of the most important applications of EAHE system. Their effectiveness was significantly improved by the use of solar air heating ducts and solar chimneys. Jakhar et al. [78] presented research on the thermal performance of Earth Air Tunnel Heat Exchanger (EATHE) coupled with a solar air-heating duct. It was shown that the heating capacity of EATHE system increased by over 5% when it was coupled with solar air heating duct, with a substantial increase in the room temperature by an average of 2.3°C higher than the base case.

Maerefat and Haghighi [89] introduced the use of a solar chimney together with EAHE systems. The conducted theoretical analysis investigated the cooling and ventilation in a solar house through combined solar chimney and an underground air channel. The findings showed that the solar chimney powered the underground cooling system during daytime, without any need to consume electricity. Although the temperature and cooling load were high, the hybrid system could provide satisfactory thermal comfort conditions, even if the solar density were low, less than 100 W/m², and the temperature were extremely high, over 50°C. In a later work [90], those same authors provided a design guideline for the application of EAHE systems coupled with solar chimney as a natural heating system, in order to meet the thermal need of flat buildings in respect to adaptive thermal comfort criteria. The proposed model was validated and showed that the overall performance of such hybrid EAHE systems depends strongly on solar radiation and air temperature. For any values of these two parameters, the optimum size of a solar chimney is 0.2m for the air gap, and 0.5m for the diameter of the heating pipe.

In an interesting work, Yu et al. [79] studied experimentally the coupling of a geothermal cooling system with earth tube and solar chimney. Room air quality, thermal comfort, thermal capabilities, and underground soil temperature analyses were performed for 43 days. The results showed that the coupled geothermal system could provide cooling to the facility in a natural operation mode without using any electricity, while the surrounding soil of the earth tube could saturate and take long to recover if the heat is over-extracted from the soil in a forced air mode. Sakhri et al. [91] have suggested an effective idea of integrating a solar chimney into a geothermal system, with the main characteristic being the east-west orientation of the facility. The air inlet would be placed in the east and be heat-insulated, in contrast to the outlet, which would be placed in the west. The air temperature difference between outlet and inlet is capable of reducing energy consumption significantly, and ultimately improving the overall performance of the hybrid system.

An alternative look at the subject of hybrid EAHE systems was provided by Benhammou et al. [92], where they proposed a typical EAHE system (buried pipe in PVC), coupled with a wind tower. A transient analytical model was developed to investigate the influence of design parameters on the performance of the EAHE. The model of the EAHE was validated against both theoretical and experimental data. The results showed that the influence of the EAHE dimensions again dominated, while the mean efficiency and the gradient of air temperature increased with (increasing) pipe length, but decreased with (increasing) pipe diameter.

The most comprehensive hybrid EAHE system proposal consisting of an electric boiler, hydrogen, wind, and PV configurations, was provided by Akhtari et al. [75]. The heat exchanger achieved improved reliability and sustainability by coupling with a hybrid renewable energy system, including wind, solar and hydrogen. The results showed that adding a hybrid RE system to an EAHE system could lead to an improvement of about 5.5% of the renewable fraction, and decreased emissions and diesel consumption by almost 48%.

# 6. Economic assessment of earth to air heat exchangers

Much of residential energy (consumed in households) is customarily used for heating, cooling, air conditioning, and production of hot water. Thus, residential energy consumption has a significant share in global energy consumption [3]. Systems that utilize RES can be a solution for the creation of buildings that are sustainable and friendly to the environment. As one such application, the ground source heat pump system is an alternative solution that can help reduce residential energy consumption [19].

The earth to air heat exchanger (EAHE) is a passive technique that is based on the usage of the underground soil temperature [18]. It consists of one or more pipes, made from metal, plastic or concrete, which are laid underground horizontally [11], [28]. EAHE systems do not require high maintenance and have a high potential for energy saving [74]. They are suitable for the heating or cooling of small and medium spaces, and have low initial and operational costs [127].

The evaluation of the performance of earth to air heat exchangers and their economic assessment is particularly interesting and has been the subject of recent literature. For instance, Chel and Tiwari [128] assessed the performance of such a system in a New Delhi (India) building and provided a life cycle cost analysis, concluding that the integration of earth to air heat exchangers has significant potential for energy saving, estimated at 10,321kWh/year, compared to 4,946kWh/year that was the energy saving potential before the integration. The annual cash inflow was estimated at $448/year and the annual cash outflow (operation and maintenance) at $240.9/year.

Ascione et al. [126] performed a technoeconomic analysis and evaluated the performance of an earth to air heat exchanger for both winter and summer in different Italian climates. It was found that such a system presented the highest efficiency for winter and summer in colder climates. In addition, if earthworks may be done inexpensively and easily, the system could be economically acceptable with simple payback estimated at 5 to 9 years.

Akhtari et al. [75] also performed a technoeconomic evaluation for an earth to air heat exchanger, coupled with a hybrid energy system including wind energy, solar energy and hydrogen. It was found that effectiveness rose by 8% during an intermittent run of the system, and energy supply was around 31.55MJ per month. If geothermal energy was added to the hybrid system, the renewable fraction could be improved by approximately 5.5%, while emissions and diesel consumption were reduced by approximately 48%.

Bisoniya et al. [129] assessed the performance of an EAHE system in the hot and dry conditions of Bhopal (India). The embodied energy of the system was estimated at 1,663.88kWh, while its maximum heating and cooling potential were found to be 191.06kWh in January and 247.25kWh in May. The energy payback period was calculated to be 1.29 years and, in its 50 years lifespan, the earned carbon credit equaled $2,837.6, while the $CO_2$ emission mitigation potential was estimated at 101.3 tons.

Li et al. [130] focused on the evaluation of a passive system that coupled EAHE with solar chimney. Such a system could provide up to 2,582W of cooling capacity, covering the cooling load of the building and maintaining thermal comfort. In another work, Li et al. [131] evaluated an EAHE system preheating fresh air in cold climate, finding that the EAHE system could lead to a temperature increase that could reach 12.4°C with an average heating coefficient of performance estimated at 29.7, meaning that that system was an effective and economic technique.

Fazlikhani et al. [27] studied the efficiency of an EAHE system in both hot and cold climates in Iran. The system could lead to improvements of the average temperature and to significant reductions in energy consumption. It was suggested that the system could be used for 294 days per year to provide energy savings of 50.1 to 63.6% in hot climates; and for 225 days per year, providing energy savings of 24.5 to 47.9% in cold climates.

Díaz-Hernández et al. [132] examined the case of an EAHE in Mexico and evaluated the system's performance under a warm humid climate. The system worked as a cooler in the morning and as a heater at night, while in the winter it could work as a heater for most of the day. Those authors also performed an analysis based on an experimental 6-month period, when the EAHE was coupled with an air conditioning system. It was estimated that the payback period was between 1 and 2 years, with the EAHE cost being $132.2 and the percentages of energy savings being 20 to 21% for most months.

Nemati et al. [133] performed an evaluation of the performance of a hybrid cooling system that combined an earth to air heat exchanger with an indirect evaporative cooler. This combination improved cooling performance, assisted in maintaining thermal comfort, and reduced energy and water consumption by approximately 62 and 45%, correspondingly. Various economic parameters were estimated, including initial investment, initial cost-saving, and annual cost savings (reduced maintenance costs, savings due to reduced electricity and natural gas demand, reduced carbon emission tax), and it was concluded that the payback time was between 5.5 and 6.4 years.

Ahmadi et al. [134] evaluated the performance of a hybrid cooling system, including an EAHE and a water spray channel, in Tehran (Iran). That work suggested that its effectiveness exceeded 100%, indicating that such a system could be used as an alternative to conventional coolers since it could provide optimal thermal comfort in the summer. Such hybrid systems are both ecofriendly and energy efficient. It was estimated that the operation cost saving could be up to 71%, with a payback period of 286 days.

## 7. Conclusions

Passive heating/cooling applications and technologies have gained ground during the last decades in order to achieve energy conservation in buildings and improve indoor and outdoor thermal comfort conditions and microclimate. Among these, earth-to-air heat exchangers have become a remarkably attractive subject for research and implementation. The present review article presented a holistic approach to the EAHE systems, including modeling the ground temperature distribution and the thermal performance of the EAHE systems; experimental evaluation; parametric studies; novel advances of hybrid technologies; and an economic assessment of system implementation. Concluding remarks are summarized below.

The ground temperature distribution at the surface and various depths below it, depend on the energy balance equation at the surface, containing the following crucial parameters: (a) solar radiation; (b) terrestrial and atmospheric radiation (long-wave); (c) convective energy between air and ground; and (d) latent heat flux due to evaporation/condensation processes. The ground temperature variation follows a sinusoidal distribution with an annual fluctuation up to 4°C and a daily one up to 0.3°C.

Many models predicting the thermal performance of EAHE systems have been developed and presented in the scientific literature. These models could be divided in three categories: (a) numerical (offering numerical solutions of complicated thermal processes including heat conduction in the soil, and convection in the soil, the tube and the air); (b) analytical (solving the heat and mass transfer equations analytically); and (c) data-driven (where modeling and prediction is achieved by training and testing the designed network with historical data). Most of the presented models were experimentally validated and had successfully predicted the thermal behavior of the systems.

A presentation of parametric studies followed, proving the importance of the main configuration parameters on system efficiency. System parameters were divided into three categories: (a) system design (pipe length and radius, burial depth, air velocity inside the tube, number of pipes, and pipe material); (b) soil types; and (c) soil surface coverage. System design parameters, especially length and burial depth, were shown to bear the most important influence on the thermal efficiency of the system.

Conducted experimental work has been mostly empirical. The reported findings are important and sufficient to highlight potential future research. As computer power and electronics evolve with unprecedented velocity and the Internet allows research findings to become rapidly accessible to the scientific community, more scientists focus on modeling. This has become obvious from the present review. However, a model requires assumptions, equations, and validation. This review has demonstrated that experimental research, which plays a key role in model validation but also sets the base for the assumptions and the range of model applicability, is mostly empirical. To support the effort of researchers working on quality model development, experimental work needs to be expanded and enhanced, so that models can drive a faster and more reliable application of heating and cooling by means of buried pipes.

While the measurement of global properties and the overall performance of a system are necessary, there are enough studies to pinpoint what the next step should be. The role of heat accumulation is an important parameter which requires more attention. Thermophysical properties of a great range of materials are poorly known and in most such applications effective properties are used. Engineering designs employing low accuracy data will inevitably provide results with a large error [135]. It is important to design appropriate laboratory experimental simulators to study the effects of individual factors or appropriate groups. Synergistic effects of natural factors should be considered, studied, and implemented. Traditional residencies and ancient building choices may provide insight as to how synergistic effects can dramatically increase the system efficiency. Finally, looking at this issue holistically, the reduction of energy consumption in a small system, such as a building, should not be examined in isolation from its environment. The energy consumption and potential pollution or ecological consequences arising from the construction and operation of such systems have to be foreseen, identified, evaluated in terms other than economic and minimalized.

**Declaration of competing interest**

The authors declare that they have no known competing financial interests or personal relationships that could have appeared to influence the work reported in this paper.

**Acknowledgments**

The authors thank Drs. T. Nadasdi and S. Sinclair for their online Spell Check Plus (https://spellcheckplus.com) that was used for proofing the entire manuscript.

**Table Captions**

Table 1: Main characteristics of considered models simulating the thermal performance of an EAHE system

Table 2: Summary of main characteristics and results of experimental studies of EAHE systems

Table 3: Parametric studies of the effect of crucial parameters on the thermal efficiency of an EAHE system

Table 4: Summary of the results of parametric studies

**Figure Captions**

Figure 1:   EAHE and main environmental and heat transfer mechanisms affecting system performance.

Figure 2:   Schematic diagram of experimental set up [114]

Figure 3:   Schematic diagram of experimental set up [116]

Figure 4:   Plan of installed EAHE system [67]

Figure 5:   Schematic diagram of experimental set up [117]

Figure 6:   Schematic diagram of experimental set up [69]

Figure 7:   Schematic diagram and experimental setup of hybrid EAHE coupled with PVT collector [80]

Figure 8:   Schematic diagram of proposed hybrid EAHE installation [83]

Figure 9:   EAHE system coupled with underground tank and solar collector for heating [87]